# Comparative survey of visual object classifiers

Hiliwi Leake Kidane
Laboratory Le2i, Universite Bourgogne - Franche-Comte,
21000 Dijon, France,

**Abstract**. *Classification of Visual Object Classes represents one of the most elaborated areas of interest in Computer Vision. It is always challenging to get one specific detector, descriptor or classifier that provides the expected object classification result. Consequently, it critical to compare the different detection, descriptor and classifier methods available and chose a single or combination of two or three to get an optimal result. In this paper, we have presented a comparative survey of different feature descriptors and classifiers. From feature descriptors, SIFT (Sparse & Dense) and HeuSIFT combination colour descriptors; From classification techniques, Support Vector Classifier, K-Nearest Neighbor, ADABOOST and fisher are covered in comparative practical implementation survey.*

## 1. INTRODUCTION

Image classification is very popular application in the image processing field. Image classification stands for identifying object(s) in given image and assigning it to a collection of objects with similar appearance, called classes [1]. Even though the objects belonging to one class have similarities regarding the type of attributes they posses, they can also be visually be notoriously different, regarding the color, size (scale), texture, design, gender (person). For humans, making the observation that a given object is present in the image is pretty straight forward and obvious, regardless of any of the previously mentioned different variations in which it comes, based on acquired knowledge. Providing this knowledge to artificial system and forcing them to acquire the human-like thinking is extremely demanding task. Additionally, problems like occlusion, scaling, pose change, clatter and many others often occur.

Object classification methodology [2], [6] consists of extracting feature descriptors [4], creating local vocabularies [5] (bag of words) using the positive training images and k-means clustering, training classifiers, extracting features from test images, creating histogram of the test feature based on the vocabularies, computing the confidence to get the true positive and false positives for the Receiver Operating Curve (ROC) curve [3].

It is always difficult to say one descriptor is good and the other descriptor is bad or one classifier is good or the other is bad without practical experimental approaches. This is due to that fact that some of them are good for some classes and the rest are good for other classes. Consequently, a comparative survey of the different object classification elements based on practical implementation is a key step to select the appropriate combination for future challenges.

In this paper, a comparative survey of different descriptors and classifiers are presented. Three different feature descriptors and four classifiers are implemented and analysed by comparing their result to similar training and test datasets.

## 2. BACKGROUND

In general block diagram of object class classification is divided into training and testing block as shown in Fig-1.

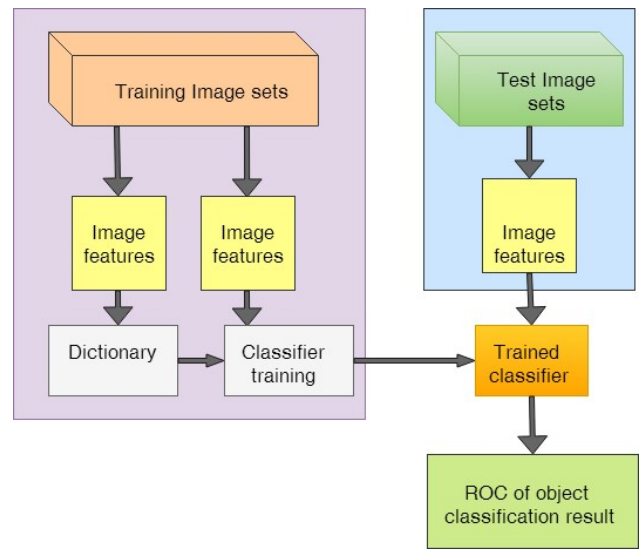

Figure 1 Block diagram of visual object classifier

In the training stage, initially a dictionary so-called "bag of words" composed of clustered positive feature descriptors of training images from each class is created. Then to train the classifiers all the positive and negative feature descriptor of the training images are extracted and created histogram length of the keywords. Finally, a classifier is trained using a set of training images so that it can be able to identify them among other words.

During the testing stage, a set of testing images are classified using the previously trained classifier, producing an output answer to the question whether the object exists in the image or not, along with the degree of certainty. The certainty is crucial measure because it represents the basis for computation of the ROC curve used for evaluating the results. The details of the descriptors, classifiers and other parameters are given below.

## 2.1. FEATURE DESCRIPTORS

Feature descriptors have been playing an important role in many computer vision problems, such as image matching and object recognition[7]. Information about object on its own is not sufficient for identification whether it is present or not in the image, due to various differences in colour, scale, viewpoint, orientation, appearance. That is why descriptors of the most salient points are extracted. They contain sufficient information for making the identification. In this paper, the following three types of feature descriptors are implemented and compared.

### 2.1.1. SIFT (vl_sift)

Prior to extracting the most salient points, a detector needs to be used. The Difference of Gaussians (DoG)[8] is the detector used for obtaining those features, followed by the use of Scale Invariant Feature Transform (SIFT) [9] with the purpose of their description. The main reason for choosing SIFT is as its name states the invariance to changes in scale, translation, rotation, local geometric distortion and furthermore to noise and different illumination. The majority of the points that SIFT extracts as salient points lie in high-contrast areas, such as object edges. As result when using the SIFT descriptor feature vectors with constant length of 128 bits are obtained. The original SIFT descriptor is also known by the name sparse SIFT. The SIFT used in this implementation is the vl_sift[10] from the VLFeat library. This is due to the fact that the vl_sift is faster than the normal SIFT implemented by David Lowe [11] with small deviation in terms of performance.

### 2.1.2. HUE SIFT (huesift)

One of the very effective and well-know SIFT descriptor extension is the so-called HUE SIFT [12]. As the name specifies it is connected with incorporating color information in the SIFT feature extraction process. One of the most distinguishable properties of this descriptor is the ability to map skin color shades for person classification necessities. HUE SIFT is scale-invariant and shift-invariant [13] (at least the SIFT component) and similarly to the hue histogram is made up by weighting each sample of the hue by its saturation. Contrary to the original SIFT descriptor the length of the extracted feature vectors is longer (165) because of the color information that is incorporated. The detail flow of this special descriptor is given in the figure 2.

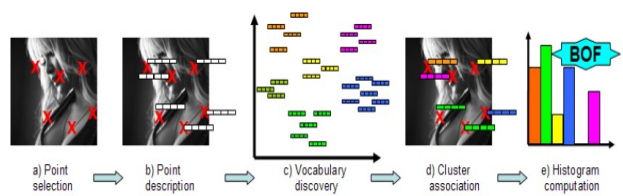

Figure 2 Steps for creating dictionary using HUE SIFT

### 2.1.3. Dense SIFT (Vl_dsift)

Dense SIFT [14] computes descriptors for densely sampled key points with same size and orientation. These key points are sampled so that the centre of the spatial bins is at integer coordinates within image boundaries. When this approach is used the number of features remains the same for images which have same size. This is one of the most characteristic differences between dense and sparse SIFT. In the case of sparse SIFT images of same size can have different sized descriptors, meaning one image may have more features, which will cause them to be weighted more. The problem can be avoided by using Dense SIFT. The features are uniformly distributed along the image providing advantage of avoiding unknown positions, but on the other side features that are not so strong can appear. In this implementation the vl_dsift [15] from the VLFeat library is used to extract the dense features. Even though the default Dense SIFT extracts the features from each pixel, in our implementation we set the pixel step to be 10 due to memory problem.

## 2.2. BAG-OF-WORDS

The Bag of Words [16], or often referred to as dictionary is built using the feature descriptors of positive training images from each class in the dataset and clustering using k-means into a matrix of length of the feature descriptor by the length of the cluster. The length of the vocabulary in the dictionary depends on the number of clusters used to create the dictionary. At the same times, the strength or performance of the dictionary depends on the number of images per class used to build the dictionary. The more the images per class used the more possibility to have distinctive vocabularies. There are several approaches to create the dictionary. They differ in the way how the positive training images are extracted and how many images per class are considered. In some cases, a unique dictionary per each class is prepared by taking features of training images a single class and clustering the features. This approach is very slow and time-consuming. Another approach is to create a single dictionary for all classes but it also follows two different ways to when taking the training images. The first way is to select the training images manually. And the second way is to select the positive training images randomly.

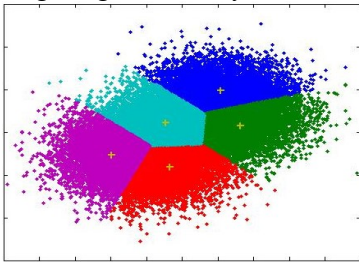

Figure 3 clustering (grouping similar attributes)

In this paper, a single dictionary is constructed for each class object for a given number of words and a number of images. To observe the effect of taking different number of words and images per class; three different lengths of clusters (50, 100, 200) and three different number of images per class (5, 15, 40) are taken. By combining the above different length of clusters and images per class, 9 different dictionaries are created. Finally using all these different dictionaries, the different types of classifiers are tested.

## 2.3. HISTOGRAM

Histogram [17] of the training image feature descriptor is created by mapping the descriptors of each positive and negative training images into the dictionary created above. The length of the histogram is equal to the length of the number of clusters. Since the dimension of the dictionary is 128xN where N is the number of clusters, the length of the histogram will be N. Every time when the descriptor of 128xK is extracted from the training images, all the K columns of the descriptor mapped into the dictionary columns of length equal to the words. When a a given column is mapped from the descriptor column to the dictionary, in the corresponding level of the histogram will be added 1.

In this comparative implementation, a MATLAB data classifier function called `knnclassify()` [18] which classifies data using the nearest-neighbor method is used. This function takes the dictionary, the descriptor and index of the cluster as input and returns the score of mapping to each index of the cluster. Finally the score of each index is added to the histogram to get the histogram of a given training class or test image as shown below. In order to have uniformity between the histogram of the training image and the test images, the histogram is always normalized.

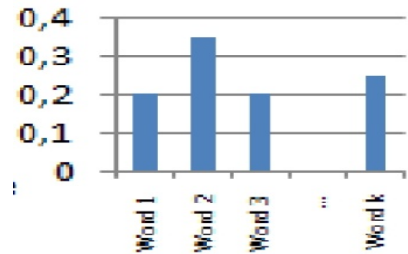

Figure 4 histogram of bag-of-words

## 2.4. CLASSIFIERS

The main task of the classifiers [19] is to divide or map the histogram created from the training images into positive and negative histogram. The classifiers take as input the histogram of a given class and the ground truth positive and negative identification of the training images. Then the classifiers map the histogram into positive and negative group so that to use as a confidence measure during the testing stage. Now we are going to get present the classifiers that were used in our implementation.

### 2.4.1. SVC(svc)

The Support Vector Classifier [20] is one of the most well-known classifiers. It is supervised method used for regression and classification. The principle of work is the following: a model is built used for making prediction whether the following (new) example is going to fall under a certain category. As input to the support vector Classifier we have set of training examples, each of whom contains a label of the category (+1,-1).

### 2.4.2. K-Nearest neighbour

The k-nearest neighbors algorithm (k-NN)[21] is a non-parametric method used for classification. It is widely known method for classifying objects based on closest training examples in the feature space is the nearest neighbour classifier. The reason why it being widely known is its simple structure, maybe even the simplest in the Classifier learning algorithm's family. The object classification is based on majority voting procedure of neighbours. The assignment to a specific class is done so that the most common class among its n-nearest neighbours is used for classification. In the most simple case, if n=1, the object is simply classified to the same class as its nearest neighbour.

### 2.4.3. Fisher

Fisher Discriminant Classifier [22] is a classification technique based on the well known fisher linear discriminant technique used to reduce dimensionality. The idea is to project the binary class in an intermediate linear space where the error of misclassification is reduced to zero thereby reducing the dimensionality.

The idea of the dimension reduction is that, it reduces the non important dimension in such a way that the retained reduced dimension represents the entire system in a robust way, but in case of classification this leads to misclassification.

### 2.4.4. AdaBoost

AdaBoost [23] is Classifier learning algorithm which initially when given a weak classifier, slightly better than random boosts the performance of that classifier to the maximum possible extent. This algorithm as most of the others is sensitive to noisy data and outliers, but less susceptible to the overfitting problem than most of the other Classifier learning algorithms. The mode of workflow is iterative, during a series of rounds, by calling a new weak classifier in each round. When the call is done a distribution of weights is being updated. This indicates the importance of examples in the data set used for classification. On each round, the weights of each incorrectly classified example are increased while, the weights of the correct classified entries is decreased. This way the classifier is forced to focus on the wrong classifications.

### 2.5. ROC CURVE

Receiver Operating Curve (ROC) [3] is the evaluation measure used for comparing the classification results. It shows the ratio between true positives (sensitivity) and false positives (specificity). In addition it returns the Area Under the Curve (AUC) where the bigger the area the better the classification system is.

## 3. IMPLEMENTATION

The detail flow of the implementation is given in Fig 5. In each step of the flow diagram, different types of entities are used. For example in the initialization of the number of clusters and images per class, three different length of number of words/cluster 50, 100, and 200 are used. At the same time for the number of images per class for creating the dictionary, three different combinations (5, 20 and 50) are used. By combining the above different of cluster lengths and images per class, 9 dictionaries were created.

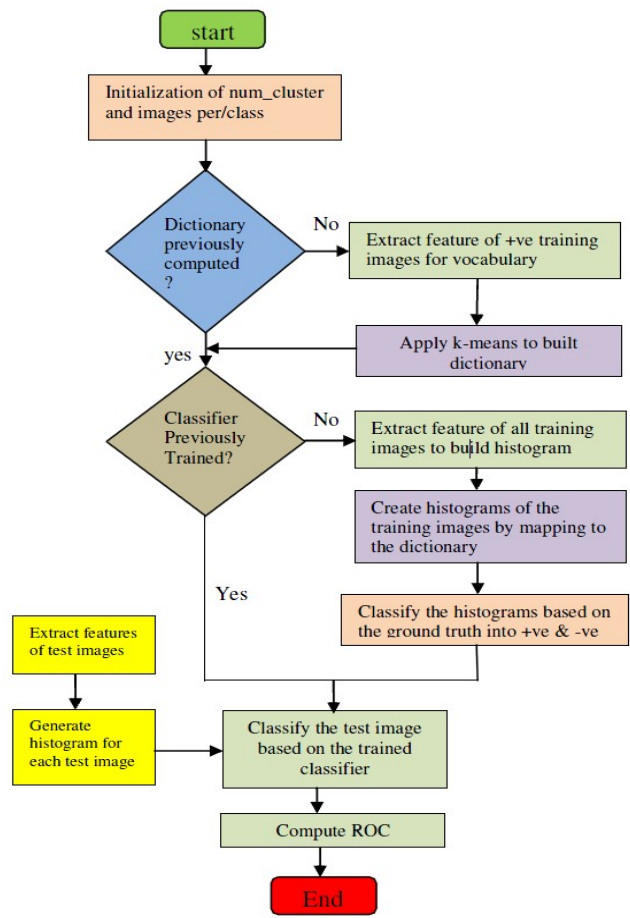

Figure 5 flow diagram of the project

In the feature extraction step three different feature descriptors are used. The first two discriptors are the sparse and dense sift from VLFeat libray which are *vl_sift* and *vl_dsift*. The third descriptor used is the *huesift* color-sift combination descriptor developed by Koen van de Sande, Intelligent System Lab Amsterdam, University of Amsterdam. Since the color descriptor software is an executable file which is ready to extract different types of color descriptors. For clustering during the creation of the dictionary, the vl_kmeans is used.

To map the training feature descriptors in to the dictionary for creating the histogram, the well known matlab mapping classifier function called *knnclassify()* is used. To train the classifier, four different classifiers from the PRTool library[24] are used. The classifiers used are SVC, FISHER, KNNC and ADABOOST.

To compute the confidence during classifying the test images, different codes for the differ classifiers is used. This is due to the fact that the mapping output of the four classifiers mentioned above is different.

## 4. RESULTS AND DISCUSSION

The final output result of the classification project is the ROC curve and the Area under curve. Even though the ROC curve gives the general pictorial true positive versus false positive rates, the area under the curve (AuC) was used as a measure of the performance of the classifier. The outputs of the classifiers are values that indicate the degree in which it belongs to the class to be selected. Some ROC curve and the AuC of the different descriptors and classifiers are given below.

1. **Vl_sift Descriptor**

The result of classification after using the vl_sift descriptor for different type of classifiers is given below. In each classifier, deferent combinations of dictionary are presented.

| Classifier : SVC(support vector Classifier) | | | | | | | |
|---|---|---|---|---|---|---|---|
| Cluster | 50 | | | 100 | | | 200 |
| Images | 5 | 20 | 50 | 5 | 20 | 50 | 5 |
| Bicycle | 0.831 | 0.825 | 0.849 | 0.841 | 0.840 | 0.850 | 0.848 |
| Bus | 0.615 | 0.667 | 0.654 | 0.610 | 0.641 | 0.606 | 0.611 |
| Car | 0.763 | 0.765 | 0.801 | 0.741 | 0.799 | 0.793 | 0.757 |
| Cat | 0.703 | 0.693 | 0.714 | 0.782 | 0.793 | 0.784 | 0.787 |
| Cow | 0.795 | 0.811 | 0.807 | 0.769 | 0.784 | 0.800 | 0.771 |
| Dog | 0.610 | 0.613 | 0.624 | 0.738 | 0.706 | 0.628 | 0.741 |
| Horse | 0.596 | 0.648 | 0.667 | 0.646 | 0.629 | 0.669 | 0.651 |
| M.bike | 0.627 | 0.603 | 0.692 | 0.606 | 0.609 | 0.642 | 0.610 |
| Person | 0.616 | 0.569 | 0.557 | 0.552 | 0.581 | 0.559 | 0.555 |
| Sheep | 0.663 | 0.728 | 0.694 | 0.682 | 0.717 | 0.710 | 0.672 |
| Mean | 0.682 | 0.692 | 0.706 | 0.697 | 0.710 | 0.704 | 0.701 |

**Table 1** AUC of SIFT +SVC

| Classifier : knn(Nearest neighbour) | | | | | | | |
|---|---|---|---|---|---|---|---|
| Cluster | 50 | | | 100 | | | 200 |
| Images | 5 | 20 | 50 | 5 | 20 | 50 | 5 |
| Bicycle | 0.816 | 0.853 | 0.863 | 0.820 | 0.860 | 0.865 | 0.822 |
| Bus | 0.696 | 0.739 | 0.714 | 0.699 | 0.741 | 0.734 | 0.699 |
| Car | 0.800 | 0.751 | 0.743 | 0.805 | 0.762 | 0.751 | 0.810 |
| Cat | 0.726 | 0.742 | 0.745 | 0.728 | 0.747 | 0.749 | 0.729 |
| Cow | 0.862 | 0.871 | 0.856 | 0.864 | 0.879 | 0.875 | 0.865 |
| Dog | 0.660 | 0.718 | 0.673 | 0.672 | 0.720 | 0.693 | 0.675 |
| Horse | 0.551 | 0.583 | 0.555 | 0.556 | 0.587 | 0.561 | 0.558 |
| M.bike | 0.513 | 0.520 | 0.551 | 0.514 | 0.527 | 0.563 | 0.516 |
| Person | 0.553 | 0.504 | 0.517 | 0.555 | 0.527 | 0.541 | 0.556 |
| Sheep | 0.747 | 0.755 | 0.772 | 0.749 | 0.763 | 0.778 | 0.750 |
| Mean | 0.693 | 0.706 | 0.700 | 0.697 | 0.713 | 0.711 | 0.698 |

**Table 2** AUC of Vl_SIFT + KNN

| Classifier : Adaboost | | | | | | |
|---|---|---|---|---|---|---|
| Cluster | 50 | | | 100 | | |
| Images | 5 | 20 | 50 | 5 | 20 | 50 |
| Bicycle | 0.793 | 0.781 | 0.760 | 0.819 | 0.762 | 0.833 |
| Bus | 0.648 | 0.669 | 0.740 | 0.608 | 0.721 | 0.620 |
| Car | 0.744 | 0.731 | 0.752 | 0.686 | 0.742 | 0.764 |
| Cat | 0.749 | 0.720 | 0.722 | 0.758 | 0.736 | 0.803 |
| Cow | 0.845 | 0.786 | 0.757 | 0.817 | 0.747 | 0.801 |
| Dog | 0.718 | 0.634 | 0.689 | 0.664 | 0.638 | 0.712 |
| Horse | 0.719 | 0.633 | 0.654 | 0.646 | 0.537 | 0.578 |
| M.bike | 0.658 | 0.711 | 0.681 | 0.641 | 0.659 | 0.704 |
| Person | 0.580 | 0.540 | 0.532 | 0.601 | 0.530 | 0.577 |
| Sheep | 0.640 | 0.659 | 0.619 | 0.725 | 0.665 | 0.729 |
| Mean | 0.710 | 0.687 | 0.691 | 0.697 | 0.674 | 0.708 |

**Table 3** AUC of Vl_SIFT + Adaboost

| Classifier : Fisher | | | | | | |
|---|---|---|---|---|---|---|
| Cluster | 50 | | | 100 | | |
| Images | 5 | 20 | 50 | 5 | 20 | 50 |
| Bicycle | 0.863 | 0.813 | 0.806 | 0.825 | 0.724 | 0.800 |
| Bus | 0.610 | 0.657 | 0.609 | 0.608 | 0.588 | 0.578 |
| Car | 0.779 | 0.764 | 0.784 | 0.714 | 0.723 | 0.757 |
| Cat | 0.693 | 0.724 | 0.697 | 0.719 | 0.695 | 0.642 |
| Cow | 0.842 | 0.852 | 0.854 | 0.767 | 0.663 | 0.835 |
| Dog | 0.651 | 0.664 | 0.670 | 0.703 | 0.563 | 0.701 |
| Horse | 0.678 | 0.677 | 0.688 | 0.523 | 0.558 | 0.580 |
| M.bike | 0.693 | 0.664 | 0.660 | 0.598 | 0.555 | 0.716 |
| Person | 0.589 | 0.591 | 0.588 | 0.587 | 0.533 | 0.556 |
| Sheep | 0.648 | 0.699 | 0.635 | 0.644 | 0.647 | 0.743 |
| Mean | 0.706 | 0.712 | 0.724 | 0.667 | 0.624 | 0.691 |

**Table 4** AUC of Vl_SIFT + Fisher

The mean result of the above 4 different classifiers is almost the same with deviation less than 0.02. In the first two classifiers SVC and Knn when the number of images and clusters for the dictionary increases there is slight improvement in the results. On the other hand, in the later classifiers Adaboost and fisher when the cluster number is increases, there is slight reduction in the performance. But this is not a general case for all classes of images.

A special observation from the K-nearest mean classifier is that its performance increase when the number of clusters increases. In the other classifiers there is no uniformity for all the classes when the number of clusters increase, but in the case of knn, when the number of clusters increases the performance also increases.

On the other hand the performance of the fisher classifier decreases when the number of clusters increases. But when the number of images per class increases, the performance of this classfier also increases.

In case of increasing images per class for the dictionary, in some classes it gives better result and in some classes lower than when small number of images are considered. Since may be due to the problem of the newly added image. Even though they are positive images, they may have some noise characteristics which affect the dictionary of the given class. So increasing images per class or number of clusters is not always guarantee for better result. It depends on the profile of images added.

The adaboost classifier is the worst for the bicycle class. In the rest three classifiers, the result of classification for the bike is greater than 0.81 but using the adaboost classifier it gives 0.793. For the cow class, the knn is the best classifier which gives up to 0.879 for number of cluster equal to 100 and images per class 20.

In terms of speed the fisher classifier is the fastest of all the four classifiers. The next fastest classifier is Knn and adaboost is the slowest. This is just comparison of the speed of the classifiers. The overall speed of the classification depends on the number of images for training and clusters. We observe that when the number of images per class and clusters are increases, the time needed for classification also increase.

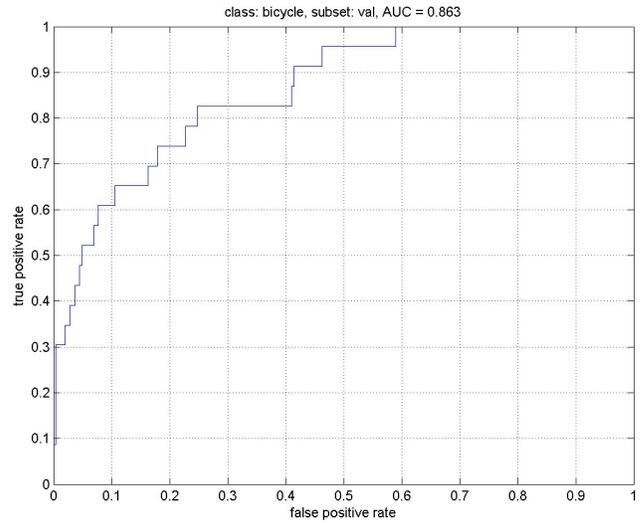

Figure 6 bicycle ROC using fisher classifier. AUC=0.863

### 2. Color descriptor (huesift)

The summary result of the color sift descriptor is given below. The length of this descriptor is 165. It comprises the local invariant characteristics of the sift descriptor and the entities of the hue colors descriptor.

| SVC(Support vector Classifier) | | | | | | |
|---|---|---|---|---|---|---|
| Cluster | 50 | | | 100 | | |
| Images | 5 | 20 | 50 | 5 | 20 | 50 |
| Bicycle | 0.822 | 0.814 | 0.801 | 0.811 | 0.820 | 0.803 |
| Bus | 0.699 | 0.750 | 0.643 | 0.622 | 0.641 | 0.606 |
| Car | 0.812 | 0.763 | 0.801 | 0.802 | 0.781 | 0.708 |
| Cat | 0.729 | 0.749 | 0.796 | 0.785 | 0.746 | 0.751 |
| Cow | 0.865 | 0.889 | 0.815 | 0.810 | 0.857 | 0.882 |
| Dog | 0.675 | 0.716 | 0.702 | 0.728 | 0.706 | 0.711 |
| Horse | 0.558 | 0.588 | 0.648 | 0.677 | 0.673 | 0.678 |
| M.bike | 0.536 | 0.558 | 0.609 | 0.693 | 0.609 | 0.678 |
| Person | 0.556 | 0.528 | 0.591 | 0.565 | 0.581 | 0.554 |
| Sheep | 0.720 | 0.737 | 0.765 | 0.714 | 0.717 | 0.710 |
| Mean | 0.698 | 0.710 | 0.715 | 0.720 | 0.713 | 0.708 |

**Table 5** AUC of HueSIFT +SVC

| Classifier : Adaboost | | | | | | |
|---|---|---|---|---|---|---|
| Cluster | 50 | | | 100 | | |
| Images | 5 | 20 | 50 | 5 | 20 | 50 |
| Bicycle | 0.782 | 0.741 | 0.755 | 0.740 | 0.752 | 0.773 |
| Bus | 0.721 | 0.728 | 0.725 | 0.610 | 0.641 | 0.606 |
| Car | 0.817 | 0.766 | 0.776 | 0.767 | 0.781 | 0.708 |
| Cat | 0.691 | 0.703 | 0.745 | 0.728 | 0.746 | 0.751 |
| Cow | 0.843 | 0.852 | 0.881 | 0.825 | 0.857 | 0.882 |
| Dog | 0.667 | 0.662 | 0.643 | 0.689 | 0.706 | 0.711 |
| Horse | 0.617 | 0.624 | 0.649 | 0.666 | 0.673 | 0.678 |
| M.bike | 0.601 | 0.655 | 0.626 | 0.606 | 0.609 | 0.675 |
| Person | 0.579 | 0.561 | 0.552 | 0.560 | 0.581 | 0.554 |
| Sheep | 0.716 | 0.678 | 0.694 | 0.682 | 0.717 | 0.710 |
| Mean | 0.704 | 0.697 | 0.705 | 0.687 | 0.706 | 0.705 |

**Table 6** AUC of HueSIFT +Adaboost

This descriptor includes the color attribute to the normal sift classifier. And in most cases to get better result of this descriptor, the classifiers should train with all possible colors of the objects in the same class. For example the color of cows is different which includes black, red, gray, white and combination of these colors. So if a red cow is missed during the training, the classifier will be weak to classify the red cows as a cow. In case of cows there might not be a problem as most of the time the back ground is green.

This descriptor is an extension of the sift descriptor to improve classification performance of some classes by adding the hue color attribute to the sift descriptor. Comparing to the normal sift used above this descriptor slightly improves the performance of the "car" class and the "cow". As it mentioned earlier, to see the advantage of using huesift over the normal sift, the number of images used for training should be as enough as all the possible colors of the objects in the class.

In this part the, result using the svc classifier and adaboost is presented. Both the classifiers when the number of images per class increases, there is slight improvement in the overall classification result.

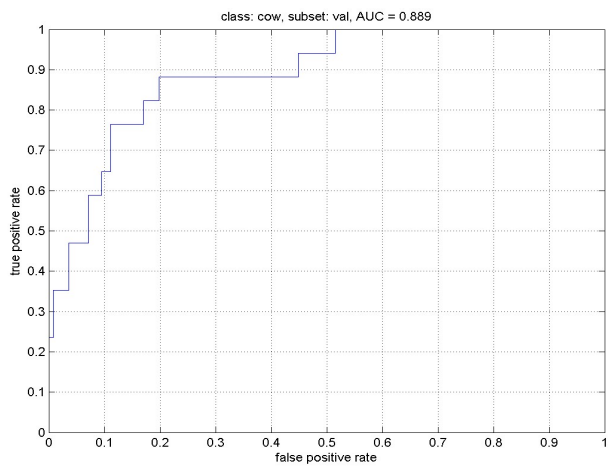

Figure 7 Cow ROC using huesift + svc . AUC=0.889

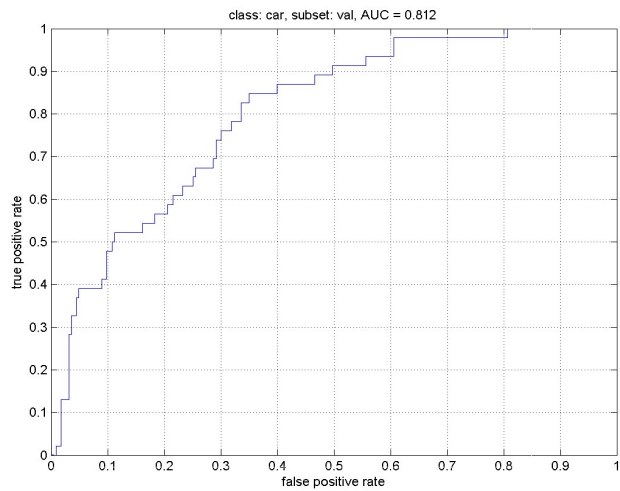

Figure 8 Car ROC using huesift + svc . AUC=0.812

### 3. Dense sift (vl_dsift)

A result of the classification using the dense sift descriptor (vl_dsift) is given below. Since the dense descriptor of a given image is a big matrix, we couldn't create the huge matrix of the positive training image descriptors to build the dictionary due to memory problem. As s result, instead of applying the full dense sift we modify the vl_dsift sampled dense descriptor after 10 pixel steps. So the following result is sampled dense sift with step between consecutive pixels equal to 10, the result couldn't be generalized as a dense sift performance. But it gives a clue about the dense sift.

| Classifier : SVC(Support vector Classifier) | | | | | | |
|---|---|---|---|---|---|---|
| Cluster | 50 | | | 100 | | |
| Images | 5 | 20 | 50 | 5 | 20 | 50 |
| Bicycle | 0.792 | 0.724 | 0.773 | 0.801 | 0.750 | 0.783 |
| Bus | 0.708 | 0.593 | 0.701 | 0.682 | 0.651 | 0.636 |
| Car | 0.875 | 0.868 | 0.872 | 0.842 | 0.791 | 0.768 |
| Cat | 0.723 | 0.633 | 0.700 | 0.785 | 0.776 | 0.781 |
| Cow | 0.877 | 0.742 | 0.856 | 0.814 | 0.837 | 0.852 |
| Dog | 0.656 | 0.666 | 0.674 | 0.689 | 0.695 | 0.711 |
| Horse | 0.604 | 0.669 | 0.671 | 0.638 | 0.673 | 0.678 |
| M.bike | 0.644 | 0.458 | 0.564 | 0.685 | 0.609 | 0.638 |
| Person | 0.566 | 0.530 | 0.542 | 0.585 | 0.551 | 0.574 |
| Sheep | 0.742 | 0.670 | 0.721 | 0.754 | 0.717 | 0.739 |
| Mean | 0.719 | 0.655 | 0.708 | 0.7275 | 0.705 | 0.716 |

**Table 7** AUC of vl_DSIFT +SVC

| Classifier : Adaboost | | | | | | |
|---|---|---|---|---|---|---|
| Cluster | 50 | | | 100 | | |
| Images | 5 | 20 | 50 | 5 | 20 | 50 |
| Bicycle | 0.843 | 0.823 | 0.804 | 0.844 | 0.835 | 0.821 |
| Bus | 0.625 | 0.656 | 0.623 | 0.644 | 0.663 | 0.679 |
| Car | 0.773 | 0.777 | 0.786 | 0.777 | 0.755 | 0.801 |
| Cat | 0.679 | 0.681 | 0.684 | 0.687 | 0.709 | 0.718 |
| Cow | 0.835 | 0.822 | 0.855 | 0.847 | 0.859 | 0.866 |
| Dog | 0.681 | 0.718 | 0.738 | 0.707 | 0.713 | 0.721 |
| Horse | 0.638 | 0.639 | 0.653 | 0.641 | 0.651 | 0.668 |
| M.bike | 0.689 | 0.668 | 0.693 | 0.677 | 0.663 | 0.616 |
| Person | 0.565 | 0.616 | 0.621 | 0.586 | 0.619 | 0.626 |
| Sheep | 0.739 | 0.758 | 0.762 | 0.755 | 0.761 | 0.772 |
| Mean | 0.707 | 0.716 | 0.723 | 0.717 | 0.725 | 0.739 |

**Table 8** AUC of Vl_DSIFT + Adaboostc

The results in Table-7 and 8 show that even using the sampled version of the dense sift, the classification output is better than the normal sift. Especially for classes Person and dog, the dense sift is better than the normal sift. The maximum classification result for the person is acquired using the dense sift descriptor.

In addition, when the number of cluster increases, the performance of this descriptor also increases. Since the number of descriptors are two much, when the cluster size increases it gives better unique representation of the descriptors by clustering in to different groups.

In general if the full dense is used the result will be much better. But there is a trade of between the performance and its drawback of computational time and memory. A sample ROC curve of the person and dog are given in Fig 9 and 10.

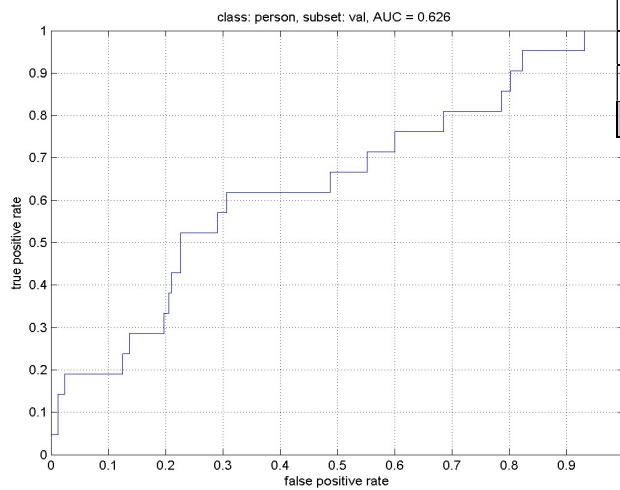

Figure 9 Perosn ROC using huesift + adaboost . AUC=0.626

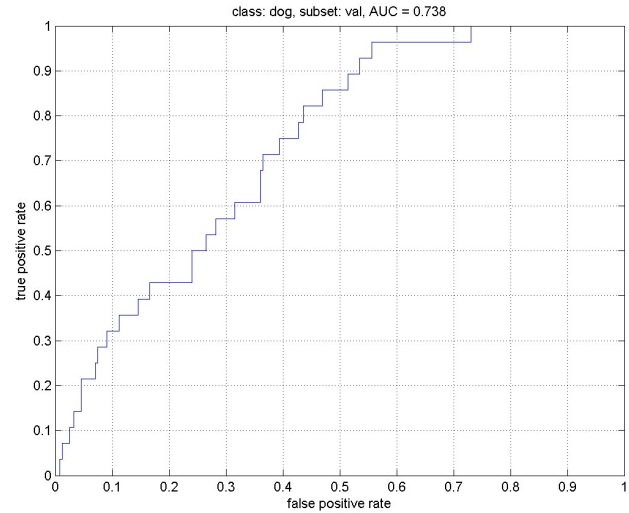

Figure 9 Dof ROC using huesift + adaboost . AUC=0.738

### 4. Comparison between descriptors

A comparison between the three descriptors using two different classifiers, cluster number equal to 50 and images per class 5 is presented below.

| | Classifier : Adaboost | | | | | |
|---|---|---|---|---|---|---|
| Cluster | SVC | | | Adaboost | | |
| Images | Vl_sift | heusift | Vl_dsift | Vl_sift | heusift | Vl_dsift |
| Bicycle | 0.831 | 0.822 | 0.792 | 0.793 | 0.782 | 0.843 |
| Bus | 0.615 | 0.699 | 0.708 | 0.648 | 0.721 | 0.625 |
| Car | 0.763 | 0.812 | 0.875 | 0.744 | 0.817 | 0.773 |
| Cat | 0.703 | 0.729 | 0.723 | 0.749 | 0.691 | 0.679 |
| Cow | 0.795 | 0.865 | 0.877 | 0.845 | 0.843 | 0.835 |
| Dog | 0.610 | 0.675 | 0.656 | 0.718 | 0.667 | 0.681 |
| Horse | 0.596 | 0.558 | 0.604 | 0.719 | 0.617 | 0.638 |
| M.bike | 0.627 | 0.516 | 0.644 | 0.658 | 0.601 | 0.689 |
| Person | 0.616 | 0.556 | 0.566 | 0.580 | 0.579 | 0.565 |
| Sheep | 0.663 | 0.750 | 0.742 | 0.640 | 0.723 | 0.739 |
| Mean | 0.682 | 0.698 | 0.719 | 0.710 | 0.704 | 0.7067 |

**Table 9** AUC of the three Descritpors

The comparison table shows that the overall heusift has some advantage than the normal sift. Sometimes the adaboost classifier gives better result when the descriptor is normal sifts than the dense sift. But this is not a general case.

Generally speaking, the huesift and dense sift are better than the normal sift. But both the huesift and dense sift have a computational problem. As the length of the huesift descriptor is 165 and the dense descriptor considers too many points per image, both of them needs higher computational memory and time.

## 5. Average Speed of descriptor and classifiers

| Desc. | Vl_sift | Heusift | Vl_dsift |
|---|---|---|---|
| Speed | fast | average | slow |

| Classif. | svc | knn | Adaboostc | Ficher |
|---|---|---|---|---|
| Speed | average | fast | average | fast |

## CONCLUSION

In this paper a comparative survey of visual object classification based on practical implementation is presented. Three feature descriptors and four classifiers are implemented and tested by creating different combinations of the feature descriptor and classifiers.

Detail analysis of the comparative survey based on the classification results for different classes is also provided. This paper will be helpful to have the general pros and cons among the different descriptors and classifiers. In addition, it will help to introduce the detail flow diagram for practical implementation of visual object classification.